\documentclass[11pt]{article}
\usepackage{graphicx,color}
\begin{document}
\begin{center}
{ \large \bf A Simulation Study on  Multicomponent Lipid Bilayer}\\[.2cm]
Srilekha Banerjee\footnote[1]{To correspond:
srilekha@boson.bose.res.in}, Jayashree Saha$^2$\\
$^1${\small S N Bose National Centre for Basic Sciences,Block JD,
Salt Lake,Kolkata 700098,India}\\
$^2${\small  Department of Physics,Visva Bharati University,Santiniketan 731235,India}
\end{center}
\begin{abstract} 
Simulation of a multicomponent lipid bilayer having a fixed percentage of
cholesterol is done to study phase transition leading to  domain formation.
The concept of random lattice has been used in  simulation to account 
for the coupling between the internal and translational  degrees of freedom 
of lipid molecules. Considering a canonical ensemble, dissimilar lipid 
molecules are allowed to exchange their positions  in the lattice subject to 
standard metropolis algorithm. The steps involved in the process  effectively 
takes into account for the movement of sphingolipids and cholesterol molecules
helping formation of cholesterol rich domains of saturated lipids as found
in natural membranes.
\end{abstract}
 
 
 
\section{Introduction}
The semipermeable plasma membrane, composed of bilayers of amphiphilic
molecules, provides the basic form and structure to the cell and allows
transportation of essential materials. These are highly flexible surfaces and
are considered to be fluid in the sense that the constituent lipid molecules
can diffuse rapidly within the membrane.
Morphologically distinct regions or domains are found on the surface membrane 
of the cells. Each of these domains are specialized for a particular function,
e.g. nutrient absorption, cell-cell communication, endocytosis etc. Lipid 
domains that include caveolae or rafts are high in cholesterol and 
sphingolipids. The lipid rafts serve as a platform
for both integral and peripheral proteins as raft lipids and associated 
proteins  diffuse together laterally on the membrane surface.\\ 
A vast literature is found on the numerous work done on lipid bilayers.
The interest has intensified in recent years
mainly because of role of sphingomyelin in lipid raft formation.  \\
Fig.1 shows a schematic picture of a lipid bilayer 
and  the interdigited hexagonal lattice  used in our model system. 
\begin{figure}[!htbp]
\centerline{\includegraphics[width=5.5in,height=5.5in]{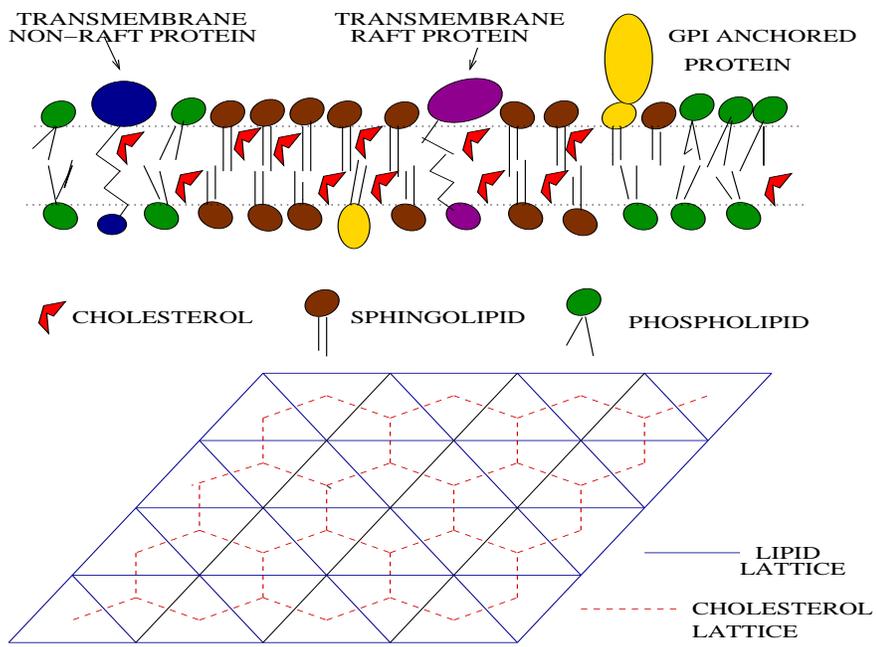}}
\caption [Fig.1:model bilayer] {Lipid-cholesterol bilayer and the model 
hexagonal lattice}
\label{bilayer}
\end{figure}
Structure and function of membrane rafts have received much attention in
recent times from many workers both theoretical and experimental
\cite{1,2,3,4,5}.\\ 
Proteins found in the membranes  either partially penetrate the hydrophobic
core or fully span the bilayer. A third possibility is getting adsorbed
to the lipid head group region. Extensive work has been done on lipid-protein
interaction by various workers \cite{6,7,8,9,10} considering different angles.
Anderson and Jacobson \cite{11} have proposed that the molecular address for
proteins targeted to lipid domains or  rafts is a lipid shell while Simons
and Ikonen \cite{12} have pointed out the importance of 
cholesterol in their formation.\\
Domain formation in model membranes composed of ternary mixtures of 
unsaturated and saturated phosphatidylcholines and cholesterol have been 
studied experimentally  by Scherfeld, Kahya and Schwille \cite{13}. 
They investigated the effect of cholesterol on mixtures of 
dioleoyl-phosphatidylcholine(DOPC) \& dipalmitoyl-phosphatidylcholine(DPPC)
and compared it to that on mixtures of DOPC \& sphingomyelin. Giant unilamellar
vesicles  prepared from ternary mixtures of various lipid compositions
were imaged by confocal fluorescence microscopy and within certain range
of sterol content, domain formation was observed. They also found evidence
of a weaker interaction of cholesterol with phosphatidylcholines than with
sphingomyelin.\\
Nielsen and co-workers \cite{14,15} have proposed a set of simple models to
study the phase equilibria in two-dimensional condensed systems of particles
where both translational and internal degrees of freedom are present and 
are coupled through microscopic interactions.
Based on the concept of random lattice they proposed two models in analogy
with spin 1/2  Ising model. In the first, the nearest neighbour particles
interact through spatially-dependent spin-spin interactions. In the second 
model, the two spin states are assigned different degeneracies, one with zero
internal (conformational) energy is non-degenerate while the other has a high 
internal energy corresponding to excitation energy associated with a
conformal change and a large degeneracy representing the large number of 
possible chain conformations having same energy. The second model is similar
to Doniach model \cite{16,17}, in which all the possible excited states of the
chain are lumped together into a single representative excited state.
The effective degeneracy of this state is taken into account. The lowest,
rigid (all trans) state of the chain is the ground state. In accordance with
Ising notation, ground state may be considered the "spin down" state and the
excited state the "spin up" state. Drawing this analogy further, the effect
of "spin" state of individual chains, relevant to its specific conformation,
on the inter-chain interaction is being incorporated in the Hamiltonian.\\
In spite of numerous work being done on lipid bilayers, the role of 
cholesterol in domain formation in a multicomponent system has not been 
investigated except for experimental studies.  We have tried to explore this 
possibility in the present work, using Monte Carlo simulation technique,  
in a model system of multicomponent lipid 
bilayer. The system comprises of a small percentage of 
sphingolipids having saturated hydrocarbon chains along with phospholipids 
having unsaturated chains forming the major component. Cholesterol molecules 
of particular concentration are introduced in the system randomly on an 
intercalated lattice. Effect of percentage variation of cholesterol
has been studied performing a series of simulations at different fixed
cholesterol concentrations. \\
The phase behaviour of lipid-cholesterol bilayer involves two distinct but
coupled order-disorder processes in terms of lipid-chain crystalline packing 
(translational degrees of freedom) and in terms of lipid-chain conformational
ordering (internal degrees of freedom). 
The concept of random lattice model \cite{14} has been used to account for the 
coupling between the two degrees of freedom  of lipid molecules, since it has 
been found experimentally that  the chain conformational and translational 
order appear/disappear simultaneously at the main transition temperature. This 
implies that the main transition involves two distinct coupled transitions.
The presence of cholesterol decouples the two ordering processes leading to a
liquid ordered ({\bf lo}) phase in which bilayers are liquid having 
translational disorder while lipid chains are conformationally ordered.\\
Cluster formation is mainly due to preferential packing of sphingolipid
and cholesterol molecules. Sphingolipid head groups occupy larger excluded 
areas in the plane of the exoplasmic leaflet than do their predominantly 
saturated hydrocarbon chains. Cholesterol molecule being hydrophobic in nature
tries to occupy the voids between associated sphingolipid chains. The hydrogen 
bonding between OH group of sterol and the amide of sphingolipid helps in the
assembly forming a separate liquid ordered phase dispersed in loosely packed
liquid disordered phase.\\
The two types of degrees of freedom are considered to be microscopically
coupled through pairwise interaction \cite{15} that include hard core repulsion
between nearest neighbours in a hexagonal lattice together with two square well 
potentials providing an approximation to the intermolecular attraction
between any two neighbouring chains.
Phase separation has been observed between {\bf so} (solid with translational 
order and collective ordering in chain conformation) and {\bf lo} (liquid, 
retaining chain conformational order) phases at low temperature and 
{\bf ld} (liquid having macroscopic disorder in both translational and chain 
conformations) and {\bf lo} phases at high temperature. 
A cholesterol molecule, not assigned  with any internal degrees of freedom, 
tries to secure ordered lipid-chain conformation in {\bf ld} phase, 
while it has a tendency to break lipid translational order in {\bf so} phase. 
Thus a cholesterol rich liquid ordered ({\bf lo}) phase is formed. The phase 
diagrams establish the coexistence of ({\bf so + lo}) phases at low
temperatures and ({\bf ld + lo}) phases at high temperatures. 
The system has been studied for different cholesterol concentrations 
but for a single canonical domain formation simulation, the cholesterol number 
is kept fixed. \\
A thermodynamic model for condensed complexes of cholesterol and phospholipid 
in a monolayer mixture was studied by Anderson and McConnell \cite{18}, while
the chemical activity of cholesterol in this context was explored  by
Radhakrishnan and McConnell \cite{19}. The phase equilibria of  ternary 
mixtures was also investigated by de Miguel and Telo da Gama \cite{20}  using 
a three dimensional continuum model of amphiphilic mixtures representing 
water, oil and surfactant. Their findings showed a region of three liquid phase 
coexistence consisting of water-rich, oil-rich and surfactant-rich phases.
Our approach is somewhat different, as has already been discussed, and the 
model ternary sytem is composed of a  
multicomponent lipid bilayer with a fixed percentage of cholesterol.
\section{Simulation Model}
The model bilayer considered in the present work is a multicomponent system 
comprising of (20\%) sphingolipids and (80\%) phospholipids. 
This composition is somewhat similar to that found in some real biomembranes 
in which the molar ratio of sphingomyclin and phospholipid  is approximately
1:4 \cite{21}. 
Lipid molecules reside at the sites of a 2 dimensional hexagonal lattice, 
defined by arrays of occupation variable  la(n2) and order parameter S(n2),
n2 = n*n being the number of lattice sites. la(i) is unity if $i^{th}$ site
is occupied by a lipid having ordered chain conformation, zero otherwise.
A cholesterol molecule is allowed to occupy only the center of triangles 
formed in the lipid lattice. The intercalated lattice so formed by joining
the centers of triangles in the main lattice specify cholesterol positions, 
which may or may not be occupied depending on cholesterol concentration.
The cholesterol lattice is defined by an array of occupation variables
lb(nc2), nc2 = 2*(n-1)*(n-1) being the number of interdigited lattice sites.
lb(i) is unity if $i^{th}$ cholesterol site is occupied, zero otherwise.   
The range for percentage cholesterol concentration $x_c$ is chosen from 
$x_c = 0$ to maximum of 20\% i.e. $x_c = 0.2$. This is justified from 
the fact that at high($\ge 25\%$) cholesterol concentration, the gel to liquid 
crystalline phase transition is eliminated and a stable liquid phase is 
produced \cite{22}.\\ 
In our model simulation, initial positions of lipid molecules, considered
to be hard core particles,  are selected 
by filling up the main lattice of dimension (n x n) randomly in accordance with
the percentage of respective lipid components, while positions of cholesterol
molecules are chosen at random on an interdigited lattice, partially occupied
only by cholesterol molecules. The system has been studied for four different 
lattice sizes namely n = 10,20,30 \& 40.\\
The acyl-chain conformational order parameter is given by
\begin{equation}
 <S> = 0.5 (3 cos^2\theta - 1) 
\end{equation}
$\theta$ being the tilt angle of lipid molecules with respect to the layer 
normal \cite{17}.
For  sphingolipids with saturated straight chains  $\theta = 0$, so the order 
parameter is 1. The unsaturated chains of phospholipids are tilted with respect
to layer normal. With increase of temperaure, thermal fluctuations destroy
long range order.  The order parameter at a given temperature t is considered 
as \cite{23,24}
\begin{equation}
 S_{ud} = S_{uo} exp(-(t-t_0)^2/12)  
\end{equation}
The reference temperature $t_0$  corresponds to fully ordered gel phase of 
unsaturated chains  having uniform molecular tilt of $\pi/6$. 
Using Eq.(1) the order parameter at $t_0$ becomes $S_{uo} = 0.625$, 
$S_{ud}$ is set to zero as it reaches a value less than a given threshold 
at a particular temperature that pertains to the main chain melting transition.
\section{The Hamiltonian}
The lateral mobility of individual lipid molecules strongly depend on
chain conformational states.
Pairwise interaction potentials couple microscopically the conformational and
translational degrees of freedom of lipid molecules.
Nearest neighbour interaction depends on the state of the particles and
inter-particle distance. It includes a hard core repulsion between nearest 
neighbours together with two square well potentials providing an approximation 
to the intermolecular forces between any two neighbouring chains \cite{14,15}.
The short range and the long range interaction potentials are given by -
$$ V^s (R) = - V^s $$
\begin{equation}
 V^l (R) = - V^l
\end{equation}
The range for short range interaction is  $ d < R \le R_0 $ while for the 
long range one  it is $ d < R \le d_{max}$,
where d is the sum of radii of two interacting particles, R the distance 
between their centers of mass, $R_0 = 1.3d_{uo}$ and $d_{max} = 1.69d_{uo}$,
$d_{uo}$ being the diameter of the phospholipid in the ordered state.\\
In the hexagonal lattice each lipid molecule has six lipid neighbours and is 
surrounded by six cholesterol sites pertaining to intercaled lattice which 
may either be occupied or not. Each cholesterol site is surrounded by three 
other filled/unfilled cholesterol sites and three lipid neighbours.
Lipid-lipid interaction is between two lipids  either both in ordered state 
or one ordered and the other in disordered state. Cholesterol interacts with 
neighbouring lipid chain which may be  either in ordered or in disordered
configuration and  also with neighbouring cholesterol molecules.  
Periodic boundary condition is considered in each case.\\
The Hamiltonian for the system  -
\begin{equation}
 H = \Sigma_i E_d*(1-la(i)) + H_{ll} + H_{lc} + H_{cc} 
\end{equation}
where $E_d$ denotes the excitation energy of the disordered state of lipid 
chains and $H_{ll}$, $H_{lc}$, $H_{cc}$ are lipid-lipid, lipid-cholesterol,
cholesterol-cholesterol interactions.
$$ H_{ll} = \Sigma_{(i<j)}V^l_{oo}S(i)S(j)la(i)la(j) +
            \Sigma_{(i<j)}V^l_{od}S(i)S(j)(1-la(i)la(j))$$
\begin{equation}
          + \Sigma_{(i<j)}V^s_{oo}S(i)S(j)la(i)la(j) +
            \Sigma_{(i<j)}V^s_{od}S(i)S(j)(1-la(i)la(j)) 
\end{equation}
$la(i)la(j)$ = 1 only if both sites have ordered chains. 
$$ H_{lc} = \Sigma_{(i<j)}V^l_{oc}S(i)la(i)lb(j) +
            \Sigma_{(i<j)}V^l_{dc}S(i)(1-la(i)lb(j))$$
\begin{equation}
         +  \Sigma_{(i<j)}V^s_{oc}S(i)la(i)lb(j) +
            \Sigma_{(i<j)}V^s_{dc}S(i)(1-la(i)lb(j)) 
\end{equation}
$la(i)lb(j)$ = 1 if $i^{th}$ lipid site is ordered and $j^{th}$
cholesterol site is occupied.\\[.2cm]
$(1-la(i))lb(j)$ = 1 if $i^{th}$ lipid site is disordered and $j^{th}$
cholesterol site is occupied.
\begin{equation}
 H_{cc} = \Sigma_{(i<j)}V^l_{cc}lb(i)lb(j) +
            \Sigma_{(i<j)}V^s_{cc}lb(i)lb(j) 
\end{equation}
$lb(i)lb(j)$ = 0 if either of the cholesterol sites is not occupied.\\
All length scales considered in the article are in angstorms.
The parameter values used for pairwise interaction potentials are 
in terms of $J_0 \equiv V^l_{oo}$, the strength
of long range interaction between chains in ordered state. The dimensionless
parameters  are $V^l_{oo} = 1.0$,
$V^s_{oo} = 1.55$, $V^l_{od} = 0.5$, $V^s_{od} = -0.5$, $V^l_{oc} = 2.45$, 
$V^s_{oc} = -1.75$, $V^l_{dc} = 0.35$, $V^s_{dc} = -0.35$, $V^l_{cc} = 0.5$,
$V^s_{cc} = -1.0$ and $E_d = 1.303$\cite{15}. \\ 
\section{Algorithm and Methodology}
We have considered four different system sizes, n = 10,20,30,40. For each of 
these, cholesterol concentration $x_c$ is varied in the range 
$0 \le x_c \le 0.2$.
Working on a particular system size and a specific cholesterol concentration,
Monte Carlo simulation was carried for temperatures in the range 
$38^0C < T \le 43^0C$. The algorithm used is as follows.\\
The initial system configuration for each simulation run pertains to fully
ordered gel phase at $38^0C$, the reference temperature $t_0$ in Eq.2. 
The phase space of the system is explored as the configuration evolves
through the following four steps, each of which is subject to Metropolis
acceptance criterion used in standard Monte Carlo simulation.\\
Accordingly in a single Monte Carlo (MC) cycle, attempts are made for - 
\begin{enumerate}
\item {\bf Transition between conformations (order-disorder) in the lipid
chains} :\\
The packing of straight chains of sphingolipids always remain ordered (S=1), 
while for phospholipids the chain conformation varies, so  the order parameter 
S may change from  $S_{uo}$ to $S_{ud}$ and vice versa.
\item {\bf Particle movement in the lipid lattice } :\\
The center of mass of the particle concerned is subject to random displacement
(dx,dy) \cite{14}, where
$$ dx = (2 \zeta_x - 1) r_{max} $$
\begin{equation}
 dy = (2 \zeta_y - 1) r_{max} 
\end{equation} 
$\zeta_x$ and $\zeta_y$ being  random numbers in the range $0 \le \zeta_{x(y)} \le 1$  and $r_{max} = 0.01*d_{uo}$. 
\item {\bf Lateral diffusion of sphingolipids} :\\
Exchange position of a sphingolipid with nearest neighbour dissimilar 
lipid molecule.  This affects interactions at both the exchanging sites.
\item {\bf Lateral diffusion of cholesterols} :\\
Move a cholesterol molecule to its nearest neighbour empty site in the 
interdigited lattice.
Each cholesterol site is surrounded by 3 other cholesterol sites and 3 lipid 
neighbours. A cholesterol molecule is moved to another neighbouring empty site 
only if the number of sphingolipid neighbours is more in the new position than
in the old one.
\end{enumerate}
The steps 1 and 2 respectively provide for changes in conformational and
translational order in lipids while steps 3 and 4 effectively takes into
account the movement of sphingolipid and cholesterol molecules. \\ 
For each of the steps mentioned above, the system lattice is scanned 
sequentially. Considering the proposed change at a lattice point, overall
impact on the system energy is taken into account using Metropolis algorithm.
If the change is favoured, new system configuration is adopted. We move
on to the next site to repeat the exercise.\\
After a sufficiently long period of equilibrium typically around 5,00000 MC
cycles, the probability distribution function $P(E,T,x_c)$ is sampled over 
next 1,000000 MC cycles .
Binning procedure is used to  store the histogram of average system energy E.
The probability distribution function at the $i^{th}$ bin is given by
\begin{equation}
 P_i(E_i,T,x_c) = \frac{1}{z} N_i e^{- E_i/T}  
\end{equation}
where\\
 $N_i$ = no. of configurations stored in the $i^{th}$ bin,\\
 T = the temperature,\\
 $x_c$ = the cholesterol concentration and \\
$ z = \Sigma_i N_i e^{- E_i/T} $ is the partition function.\\
The transition temperatures were found from peaks of specific heat plots
against temperature. It may be noted that since we have considered the 
multicomponent system as a whole in our simulation, there was no need 
of taking the melting temperatures of individual components.\\
The histogram of energy stored can be used to generate data at a temperature
close to the actual simulation temperature using reweighting technique of
Ferrenberg and Swendsen \cite{25,26,27}. The probability distribution function
at a temperature $T'$ is computed as -
\begin{equation}
 P'_i(E_i,T,x_c) = \frac{P_i(E_i,T,x_c) e^{E_i(\frac{1}{T'} - \frac{1}{T})}}
		  {\Sigma_i P_i(E_i,T,x_c) e^{E_i(\frac{1}{T'} - \frac{1}{T})}} 
\end{equation}
The free energy corresponding to the $i^{th}$ energy bin is given by 
\begin{equation}
F_i(E_i,T,x_c) = -log(P_i(E_i,T,x_c))
\end{equation}

\section{Results and Discussion}
The simulation has  been carried out for each of the four system sizes, 
mentioned before,  using a range of values of percentage cholesterol 
concentration $x_c$, starting from $x_c = 0$ (pure system) to $x_c  = 0.2$. 
It is found that at very low cholesterol concentration, only the main chain 
melting transition {\bf so-ld} is present, {\bf so} being the gel phase and 
{\bf ld} corresponds to liquid crystalline $L_\alpha$ phase having macroscopic 
disorder in both translational and chain conformations. At intermediate 
cholesterol concentration $(0.05 \le x_c \le 0.15)$, coexisting  phases 
are observed. 
Fig.2 shows the phase diagram as a function of cholesterol concentration $x_c$
and reduced temperature $T/T_M$, $T_M$  being the transition 
temperature at $x_c = 0$. In the absence of cholesterol, there is only 
gel-fluid transition i.e. solid-ordered to liquid-disordered ({\bf so-ld}) 
phase giving $T/T_M = 1$.
For higher cholesterol concentrations, below the main chain melting transition,
a transition from fully ordered ({\bf so}) to mixed phase ({\bf so+lo}) is 
observed and above the main transition at higher temperature a fully disordered
phase is found i.e. a transition from mixed ({\bf ld+lo}) to disordered liquid
phase ({\bf ld}) occurs. The phase diagram shows these  transitions for 
$x_c \le 0.15$.
\begin{figure}[!htbp]
\centerline{\includegraphics[width=5.5in,height=5.5in]{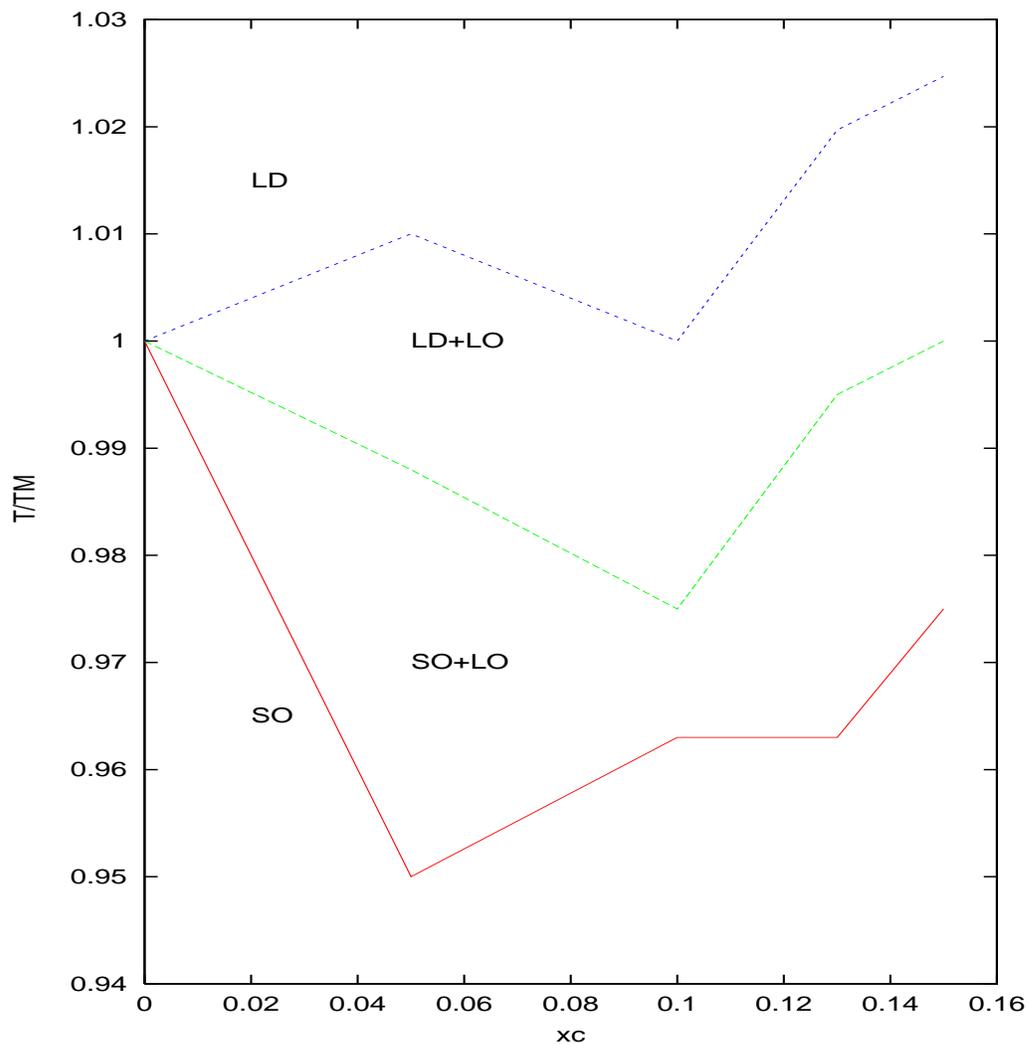}}
\caption {Phase diagram for the lipid-cholesterol model as a function of 
cholesterol concentration $x_c$ and reduced temperature $T/T_M$, $T_M$ being
the main transition temperature at $x_c = 0$. The different phases labeled
are {\bf so}, solid-ordered (gel); {\bf ld}, liquid-disordered (fluid) and 
{\bf lo}, liquid-ordered, the first letter refers to lateral order of the 
phase while the second the conformational order.}
\end{figure}
At high cholesterol concentration $(x_c \approx 0.25)$ 
gel-fluid transition is completely eliminated and a stable liquid phase with
relatively high orientational order {\bf lo} is produced.\\ 
The reweighting technique has been adapted to calculate the distribution 
function $P(E,T,x_c)$ using Eq.(10) for a range of temperature very close to 
the transition region in each of the three cases i.e. so to so+lo, so+lo to 
ld+lo and ld+lo to ld transitions. From this the free energy $F(E,T,x_c)$ is 
calculated by Eq.(11). In the $F(E,T,x_c)$ vs. E plot, pronounced double minima 
corresponding to two coexisting phases is found at each transition point.
Fig.3 shows the reweighting plots at the main transition temperature, $\Delta T
= \pm 0.005$ for $x_c = 0.15$ and n = 40. 
The middle curve corresponds to the finite-size equilibrium 
transition temperature.\\
\begin{figure}[!htbp]
\centerline{\includegraphics[width=5.5in,height=5.5in]{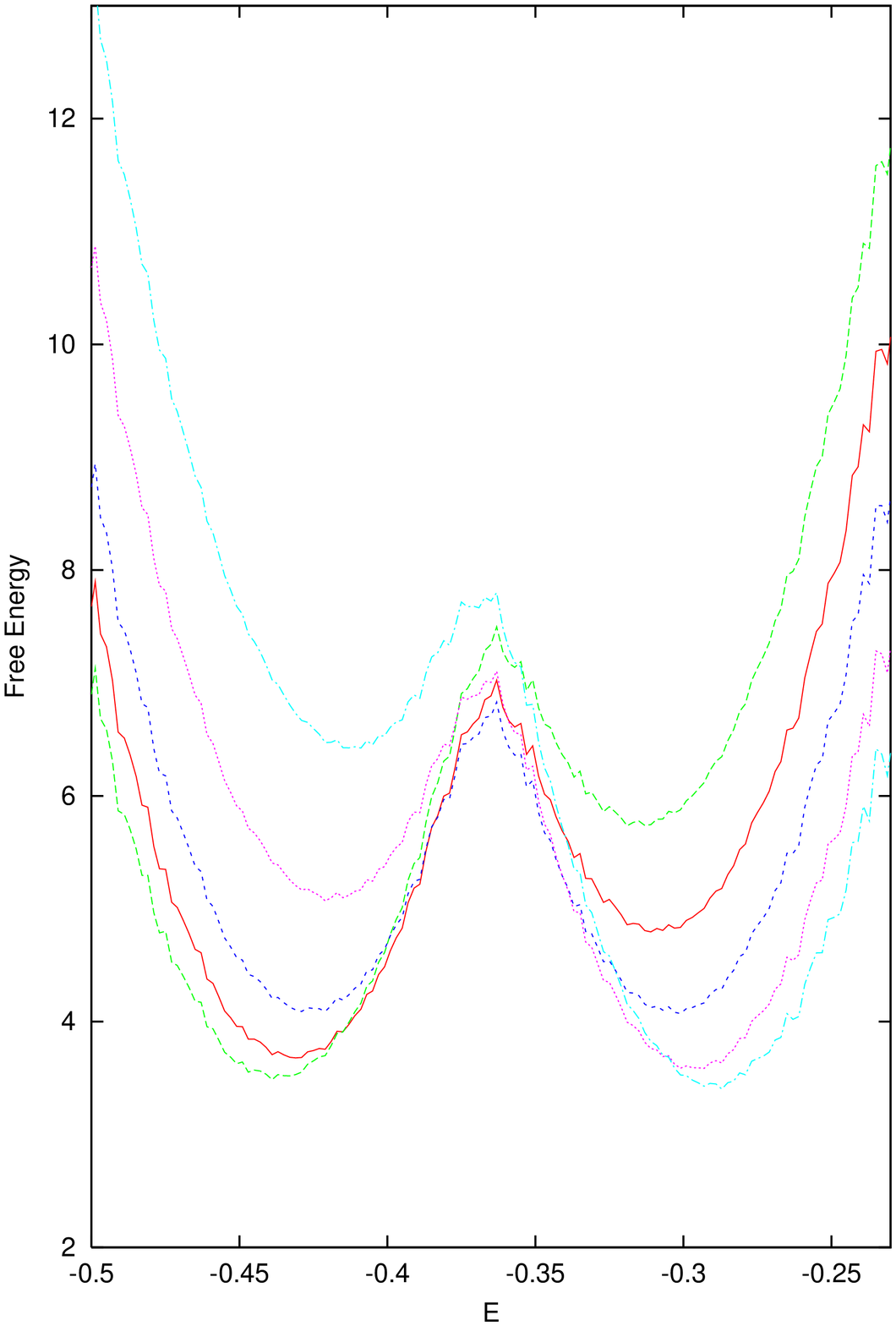}}
\caption {Free energy $F(T,x_c)$ plots for n = 40 for temperatures close
to the equilibrium transition temperature. Using reweighting technique,
extrapolations are done for temperature difference of $\triangle T = \pm0.005$.}
\end{figure}
Fig.4 shows the free energy plot at the main transition temperature 
$T_c = 40.5^0C$ for size-4 (n=40) at $x_c = 0.15$.
\begin{figure}[!htbp]
\centerline{\includegraphics[width=5.5in,height=5.5in]{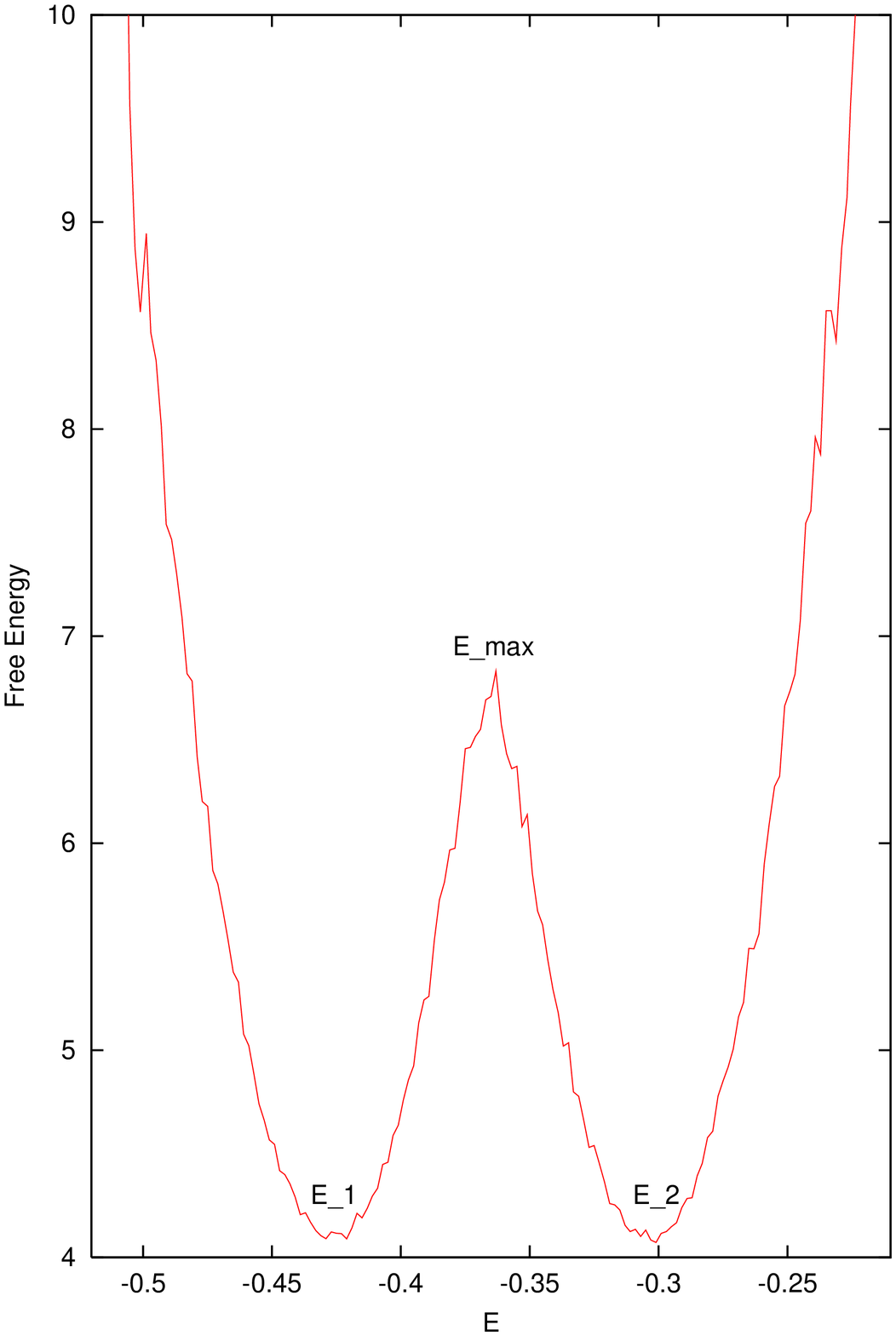}}
\caption {Free energy plot at the main transition for n= 40 and $x_c = 0.15$.
Double well structure confirm the presence of coexisting phases.} 
\end{figure}
The two coexisting phases at $E = E_1$ and $E = E_2$ are separated by a barrier 
$\Delta F(T,x_c)$, with a maximum at $E_{max}$ corresponding to interface 
between the two phases. The height of the barrier measures the interfacial
free energy between the two coexisting phases and is given by 
\begin{equation}
\Delta F(T,x_c) = F(E_{max},T,x_c) - F(E_1,T,x_c)
\end{equation}
It is found that the double well structure is more prominent with larger
system. So the barrier height $\triangle F_L(T,x_c)$ increases as the system 
size L(= nxn) increases. This can be seen in Fig.5. 
The barrier height is calculated by Eq.12 from the free energy plots at 
main transition for different system sizes. For $x_c = 0.15$, barrier heights 
are plotted against four lattice sizes used in the simulation.\\
\begin{figure}[!htbp]
\centerline{\includegraphics[width=5.5in,height=5.5in]{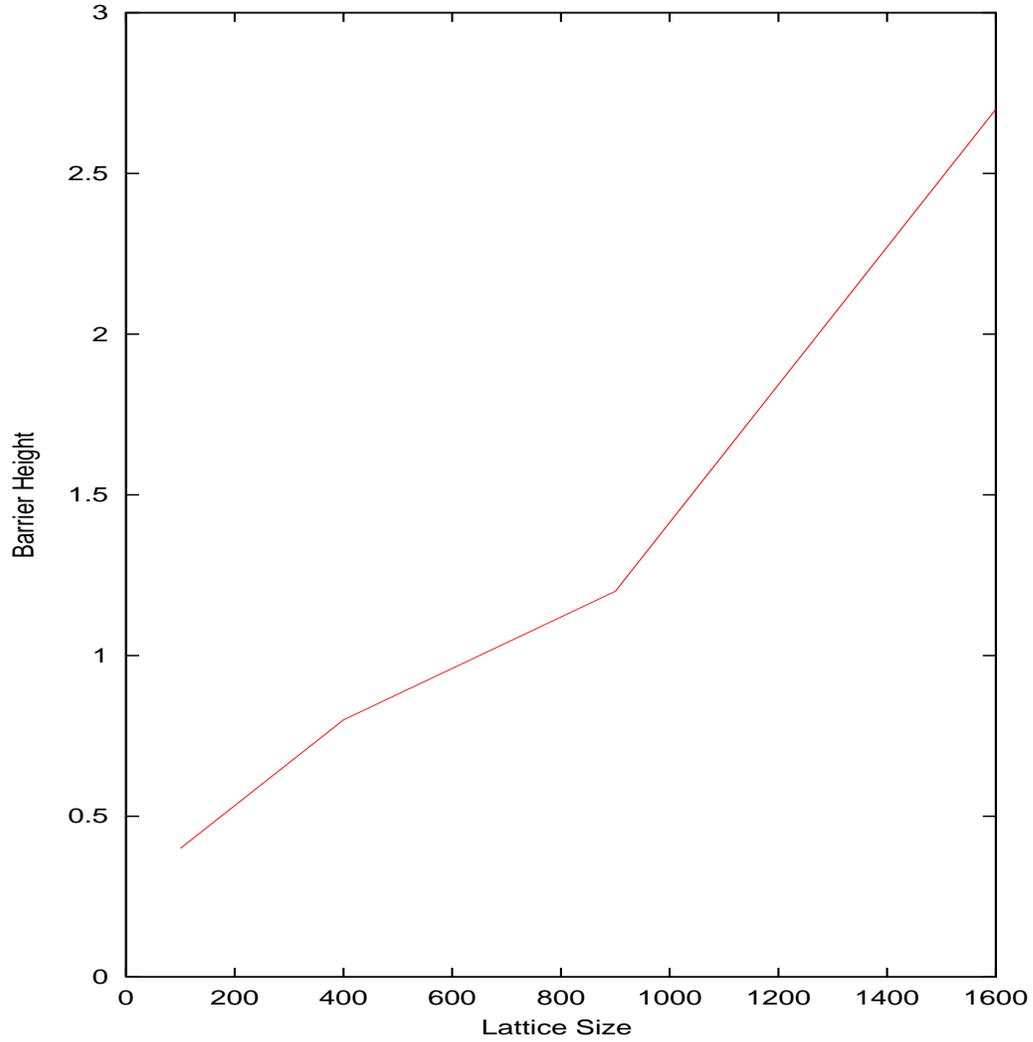}}
\caption {Barrier heights as a function of lattice size L. The plot shows 
increase with system size in the 'interfacial energy' $\triangle F_L(x_c)$,  
defined as the height of the maximum relative to the two minima of spectral
free energy function} 
\end{figure}
At low concentrations, cholesterol molecules predominantly influence the
conformational degrees of freedom of lipid molecules and tend to promote
the domain formation thus enhancing the dynamic membrane heterogeneity.
This effect is illustrated in Fig. 6 for a range of cholesterol concentration.
The cholesterol molecules tend to accumulate at the domain interfaces as can 
be seen in the plots.\\  
\begin{figure}[!htbp]
{\includegraphics[width=5.5in,height=5.5in]{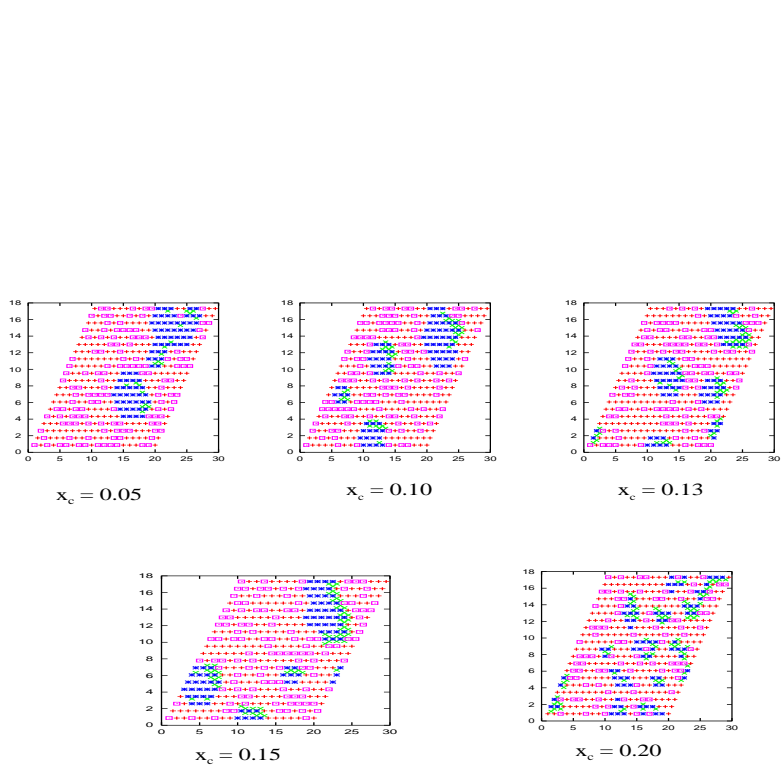}}
\caption {Simulation results showing formation of clusters of sphingolipids
(\textcolor[rgb]{0,0,1}{*}) \& cholesterol(\textcolor[rgb]{0,1,0}{x}) at
different cholesterol concentrations. The membrane configurations demonstrate
formation of domains in the cholesterol range $0 < x_c \le 0.2$.}
\end{figure} 
\section{Concluding Remarks}
In the present work the phase coexistence has been located by computing the 
free energy using distribution functions or histograms of internal energy,
reweighting of distribution functions very close to coexistence or phase 
transition and then a subsequent analysis of size dependence by finite-size
scaling theory. In the process we have explored the role of cholesterol
at different concentration in the formation of domains. It establishes 
experimentally observed facts to a large extent.\\
The lipid bilayers that are locally ordered and dynamically organized due
to fluctuations and cooperative modes controlled by the underlying phase
equilibria are sensitive to molecular agents, active at lipid domain 
interfaces, that can lower the interfacial tension. Cholesterol is one such 
agent which can alter the lateral structure of the lipid bilayer. In order
to understand the full phase diagram of lipid-cholesterol mixtures, 
cholesterol's coupling to the translational degrees of freedom has to be
taken into account. This has been done in the present work by using a model
similar to Doniach model on a random lattice. The cholesterol with its
smooth sterol skeleton, is able to break the lipid translational order
and at the same time stabilizes the ordered lipid-chain conformation. Hence 
the effect of cholesterol is to decouple the translational and conformational
degrees of freedom.\\ 
As discussed by Scherfeld and coworkers\cite{13}, the intermolecular 
interactions between sphingolipid and cholesterol is further strengthened
by the hydrogen bonds induced by the amide group at the polar-apolar interface,
which can act both as hydrogen bond-donating and -accepting group. This makes
the sphingolipid-cholesterol interaction stronger than  the 
phospholipid-cholesterol interaction. As a result, cholesterol intercalates more
tightly in sphingomyelin bilayers than in glycerophospholipid bilayers.\\
An important feature of our work is that we have used a 
multicomponent lipid system
having lipids with saturated as well as unsaturated hydrocarbon chains. This 
is essential for raft formation. Sphingolipids containing long largely 
saturated acyl chains readily pack tightly together making enough room in 
between for cholesterol occupation. Secondly, though the concept of random 
lattice has been used to ensure coupling of translational and internal
degrees of freedom of the lipid molecules, we have worked throughout with 
constant lattice size. This is because we are interested only in the formation
of clusters of sphingolipids and cholesterol molecules. The exchange of 
dissimilar lipid sites  and movement of cholesterol molecules in the
intercalated lattice subject to conditions mentioned earlier  suffices our 
requirement. \\
We further need to focus  on the affinity of certain kinds of proteins 
towards these condensed-complexes of sphingolipids and cholesterol molecules.
The molecular packing created by cholesterol and sphingolipids, as has been 
found in the present simulation, may as well act as the first step to 
simulate formation of rafts. Considering all these aspects, we intend to study 
a multicomponent bilayer, comprising of phospholipids, 
sphingolipids, cholesterols and proteins in appropriate percentages, in aqueous 
environment to understand the mechanism of raft formation.

\end{document}